# Probing the structural stability of R-phycocyanin under pressure

Simeon Minic[a,]*, Luka Velickovic[a], Burkhard Annighöfer[b], Aurélien Thureau[c], Nikola Gligorijevic[d], Zorana Jovanovic[a,b], Annie Brûlet[b], Sophie Combet[b,]*

[a]University of Belgrade - Faculty of Chemistry, Center of Excellence for Molecular Food Sciences & Department of Biochemistry, Studentski trg 12-16, 11000 Belgrade, Serbia

[b]Laboratoire Léon-Brillouin (LLB), UMR12 CEA, CNRS, Université Paris-Saclay, CEA-Saclay, F-91191 Gif-sur-Yvette CEDEX, France

[c]SWING beamline, Synchrotron SOLEIL, Saint-Aubin BP 48, F-91192 Gif-sur-Yvette, France

[d]University of Belgrade - Institute of Chemistry, Technology, and Metallurgy, National Institute of the Republic of Serbia, Department of Chemistry, Studentski trg 12-16, 11000 Belgrade, Serbia

*Correspondance to:

sminic@chem.bg.ac.rs; Tel : +381113282111

sophie.combet@cea.fr; Tel : +33169086720

**Running title**: R-phycocyanin under pressure
**Total number of manuscript pages**: 36 (8 figures, 1 table)
**Total number of Supplementary material pages:** 5 (3 figures).
**Supplementary material file name:** Minic et al_Protein Science_Probing the pressure stability of R-PC_Supp Mat

# ABSTRACT

The red macroalgae *Porphyra*, commonly known as Nori, is widely used as food around the world due to its high nutrient content, including the significant abundance of coloured phycobiliproteins (PBPs). Among these, R-phycocyanin (R-PC) stands out for its vibrant purple colour and numerous bioactive properties, making it a valuable protein for the food industry. However, R-PC's limited thermal stability necessitates alternative processing methods to preserve its colour and bioactive properties.

Our study aimed to investigate the in-situ stability of oligomeric R-PC under high pressure (HP) conditions (up to 4,000 bar) using a combination of absorption, fluorescence, and small-angle X-ray scattering (SAXS) techniques. The unfolding of R-PC is a multiphase process. Initially, low pressure induces conformational changes in the R-PC oligomeric form (trimers). As pressure increases above 1,600 bar, these trimers dissociate into monomers, and at pressures above 3,000 bar, the subunits begin to unfold. When returned to atmospheric pressure, R-PC partially refolds, retaining 50% of its original colour absorbance. In contrast, heat treatment causes irreversible and detrimental effects on R-PC colour, highlighting the advantages of HP treatment in preserving both the colour and bioactive properties of R-PC compared to heat treatment.

**SIGNIFICANCE:** HP is a powerful probe that reveals intermediate states of proteins through subtle structural changes not accessible by other denaturation methods. By combining HP-small-angle-X-ray scattering with HP-absorption and fluorescence spectroscopy, we elucidate the multiphase unfolding process of R-phycocyanin. This process includes: 1) conformational changes, 2) oligomer dissociation at moderate pressures, and 3) monomer unfolding. Our approach provides new opportunities for the structural determination of protein intermediates and oligomers using HP.

1. INTRODUCTION

Alternative proteins from algae represent sustainable and promising solutions to replace proteins from animal origin, resulting in positive environmental and health impacts (Geada et al. 2021). Red algae *Porphyra*, commonly called Nori, is rich in proteins and has high annual production (millions of tons), making it a promising alternative protein source (Rawiwan et al. 2022; Velickovic et al. 2023). The major proteins of *Porphyra*, R-phycoerythrin (R-PE) and R-phycocyanin (R-PC), belong to the phycobiliprotein (PBP) group, and their abundance could go up to 2-3% of algae dry weight (Cao et al. 2016). These proteins are highly water soluble and bind characteristic tetrapyrrole chromophores with vivid colours and strong bioactive properties. This increases R-PE and R-PC potential for applications as alternative proteins and gives them a significant advantage over other non-chromophore-containing proteins (Simovic et al. 2022; Velickovic et al. 2023). Compared to R-PE, which has been thoroughly studied, R-PC from *Porphyra* sp. remains underexplored (Velickovic et al. 2023).

Phycoerythrobilin (PEB) and phycocyanobilin (PCB) are covalently attached to R-PC *via* thioether bonds, imparting this protein with its intense purple colour. R-PC consists of two subunits: α, with a molecular weight of 18-20 kDa, and β, with a molecular weight of 19-21 kDa (Jiang et al. 2001; Velickovic et al. 2023). Each subunit, α and β, contains one PCB chromophore, while the β subunit additionally contains one PEB chromophore.

R-PC has limited thermal stability, which significantly hampers its potential for applications as a food colourant (Velickovic et al. 2023). It is sensitive to high temperatures, and thermal treatment at 60°C irreversibly removes its vivid purple colour (Velickovic et al. 2023). Therefore, alternative food processing methods containing *Porphyra* PBPs are required to preserve their colour. High-pressure processing (HPP) of food is environmentally friendly and enables better preservation of nutrients, colours, and flavours, providing a promising solution for treating PBP-containing food. Indeed, HP treatment of R-PE from *Porphyra* (Simovic et al. 2022) and of C-phycocyanin from *Spirulina* (Shkolnikov Lozober et al. 2023; Zhang et al. 2021) better preserves their colour compared to thermal treatment. However, although it is a milder approach in food treatment compared to temperature, HP could also induce protein denaturation and, consequently, partial colour loss in PBPs (Faieta et al. 2021; Simovic et al. 2022; Zhang et al. 2021). It is therefore essential to elucidate the unfolding mechanism of PBPs under HP to develop new approaches that enhance their stability under such condition.

Despite the importance of HP in the food industry, fundamental studies on the effects of pressure, especially concerning *in situ* HP-induced protein structural changes, remain scarce and largely insufficient to establish trends and understand the involved mechanisms (Minic et al. 2020; Minic et al. 2022). Pressure delivers a unique set of conformational states with specific structural and thermodynamic properties, enhancing our understanding of protein folding and unfolding mechanisms (Rouget et al. 2010). It is well known that HP disrupts the native structure of proteins by decreasing the protein volume upon denaturation (Minic et al. 2020). Pressure mostly unfolds proteins because hydrophobic cavities present in the folded state are eliminated in the unfolded states (Roche et al. 2012). Internal cavities are thus critical structural features of proteins and at the origin of fluctuations between different conformational states (Minic et al. 2020; Minic et al. 2023). The central hole in R-PC oligomers (Fig. 1, A-B), as well as many binding sites of phycobilin chromophores (nine molecules *per* trimer of R-PC), make this protein a relevant model system for studying the roles of cavities in the stability of oligomeric proteins under pressure (Jiang et al. 2001).

In the present work, we studied the effects of HP treatment on the *in situ* and *ex-situ* structural stability of R-PC in *Porphyra haitanensis* red macroalgae. Our goal was to elucidate the mechanisms of R-PC unfolding. We used *in situ* HP-small-angle X-ray scattering (SAXS) to probe the three-dimensional conformational changes of the R-PC protein. This was combined with *in situ* HP-visible (VIS) absorption and *in situ* HP-fluorescence to monitor changes in the local environment of its tetrapyrrole chromophores, phycocyanobilin and phycoerythrobilin.

## 2. RESULTS

### 2.1 Structural changes of R-PC under HP probed by SAXS

SAXS spectra ($I(Q)$) of R-PC at pH 5.7 and near atmospheric pressure (30 bar) have a characteristic shape with a pronounced shoulder at $Q = 0.1$ Å$^{-1}$ (Figs. 1C and 2A). The Kratky representation ($Q^2 I(Q)$ *versus* $Q$), which is a valuable indicator of the protein folding state, reveals the presence of two peaks at $Q$ equal to 0.04 and 0.1 Å$^{-1}$. The presence of a well-defined peak at 0.04 Å$^{-1}$ demonstrates that the protein is rather compact and well-folded (Fig. 2B). The existence of a specific shoulder in the SAXS curves (Fig. 2A) and the well-pronounced peak at 0.1 Å$^{-1}$ in Kratky plot (Fig. 2B) indicates the presence of a central hole in the characteristic oligomeric structure of PBPs (Velickovic et al. 2023).

Furthermore, we checked that the native R-PC at pH 5.7 is trimeric (($\alpha\beta$)$_3$) by fitting the SAXS data with the theoretical curve (Fig. 1C) obtained from the R-PC crystal structure ~~(PDB:1F99) using CRYSOL software (**Franke et al. 2017**)~~. The good correlation between the R-PC crystal structure and its structure in solution was confirmed by *ab initio* GASBOR modelling of the SAXS curve (Fig. 1B) ~~using GASBOR software (**Svergun et al. 2001**)~~. The obtained shape of R-PC closely resembles the protein crystal structure (Fig. 1A).

Increasing the pressure up to 600 bar (60 MPa) induces an immediate decrease in the scattering intensity (Fig. 2A) and, consequently, of the $I(0)$ values (Fig. 2C). $I(0)$, which reflects the volume of the protein and the contrast value (with hydration-dependent scattering length densities (SLDs)), decreases smoothly in the range from 500 to 1,500 bar. In comparison, a steep decrease is observed above 2,000 bar (Fig. 2C). Meanwhile, in Kratky representation, two-peak patterns are observed, which are well preserved up to 1,600-1,800 bar (Fig. 2B). In addition, the radius of gyration ($R_g$) (calculated by Guinier approximation), which is an indicator of protein compactness and size, is mostly constant (between 41 and 42 Å) up to 2,000 bar (Fig. 2D). All these results indicate that the compact trimeric form of R-PC is preserved at least up to 1,600 bar. However, the structure is more compressed compared to its form at near-atmospheric pressure (30 bar).

~~In order to confirm this hypothesis, we performed both (*i*) SasView analysis by comparing the experimental SAXS curves at different pressures with the model of a hollow cylinder (the shape that mainly reproduces the one of native R-PC) (Fig. 3) and (*ii*) *ab initio* modelling of experimental data using GASBOR software (ATSAS) (Fig. 4).~~

Similarly to the *I*(0) behaviour (Fig. 2C), the hollow cylinder analysis reveals a significant decrease (30%) of the cylinder length up to 600 bar, followed by almost constant values up to 1,600 bar and an additional substantial drop above this pressure value (Fig. 3A). The pressure increase also induces the decrease of the cylinder inner radius (central hole), accompanied by the increase of the cylinder thickness (Fig. 3B). On another hand, the total radius of the cylinder is almost constant, mainly up to 1,800 bar (Fig. 3A). We also observed that $R_g$ values calculated from the hollow cylinder parameters (using Eq. 5, see Section 5.6) are mostly constant up to 2,000 bar (37-38 Å, data not shown), which is in the reasonable agreement with the values obtained by Guinier approximation.

GASBOR a*b initio* modelling matches the experimental curves well, as shown in the Kratky representations and corresponding error distribution plots (Fig. S1). The obtained models confirm that the trimeric R-PC structure (Fig. 4A) is preserved up to 1,600 bar. However, an increase of pressure, even from 30 to 200 bar, significantly reduces the modelled thickness (Fig. 4B), confirming the results obtained from the fits with the hollow cylinder model (Fig. 3B). GASBOR modelling also demonstrates that the pressure increase from 30 to 200 bar induces a substantial shrinking of the central hole. However, interestingly, the total radius of the R-PC trimer does not change significantly at lower pressure (< 1800 bar), as confirmed by both hollow cylinder analytical fitting (Fig. 3A) and GASBOR modelling (Fig. 4B). Finally, *Ab initio* modelling demonstrates that R-PC at 1,600 bar has a triangular shape, which is unrealistic, indicating that the partial dissociation of trimers occurs at that pressure. Indeed, at 1,800 bar, GASBOR modelling cannot recover the trimeric shape of R-PC, but rather a horseshoe shape, which may be ascribed to R-PC dimers (Fig. 4, A-B).

Pressure above 2,000 bar induces significant changes in Kratky plots. The peaks become less pronounced and vanish (Fig. 2B) and the length of the hollow cylinder is substantially reduced (Fig. 3A) (despite the bad quality of the fit with a hollow cylinder model above this pressure, as shown in the plots of error distribution (Fig. S2)). All these results nevertheless confirm that the trimeric structure of R-PC is lost above 2,000 bar.

Additionally, $R_g$ increases above 2,000 bar and reaches a plateau at 3,200 bar (Fig. 2D), indicating the existence of less compact structures related to protein unfolding. A similar finding is found by GASBOR analysis at 2,600 bar, which gives models resembling the extended (αβ) monomers (Fig. 4B). Further pressure increase induces additional unfolding of R-PC, resulting in more elongated structures observed at 3,000 bar, indicating protein unfolding (Fig. 4B). However, $I(0)$ values decrease only by 30%, while increasing the pressure from 2,600 to 4,000 bar, indicating that the R-PC subunit association persists at least partly, even at denaturating pressures. Pressures above 3,200 bar induce additional elongation of R-PC, as deduced from the increase of $R_g$ values (Fig. 2D).

**2.2 Structural changes of R-PC under HP probed by absorption and fluorescence**

The VIS absorption spectrum of R-PC exhibits two peaks, at 552 and 617 nm (Fig. 5A), originating from one PEB and two PCB chromophores *per* αβ monomer, respectively (Velickovic et al. 2023). An increase of pressure up to 1,400 bar does not significantly change the absorption intensity of both peaks (Fig. 5, B-C). Hence, it seems that the protein conformational changes (*i.e.* compression of R-PC trimeric structure) observed by SAXS up to 1,600 bar do not change the absorption intensity of R-PC chromophores. Further pressurisation of R-PC substantially decreases the intensity of the chromophore absorbance, the peak at 617 nm, attributed to the PCB chromophore, being much more pressure-sensitive than the peak at 552 nm, attributed to the PEB (Fig. 5, B-C). Therefore, a significant perturbation within the R-PC structure, such as trimer dissociation and unfolding, is required to decrease protein absorption substantially. The transition curves, obtained by monitoring the pressure dependence of normalised absorption (Eq. 1, see Section 5.3) (Fig. 5, B-C), reflect both the dissociation and unfolding of R-PC. These curves are well fitted by Eq. 2 (Section 5.3), and we obtained the thermodynamics parameters of R-PC dissociation/unfolding: apparent volume change ($\Delta V$), Gibbs free energy ($\Delta G$ at 1 bar) and half-transition point ($P_{1/2}$) (Table 1). The dependence of both absorption maxima with pressure gives similar thermodynamic parameters, indicating that both subunits have similar pressure stability.

While the pressure dependence of absorption intensity gives simple one-transition curves, monitoring the absorption wavelength under increasing pressure reveals the complex behaviour of R-PC. The absorption peak from the PCB chromophore undergoes a continuous red shift from 617 to 622 nm up to 2,600-2,800 bar (Fig. 5D). This shift is less pronounced at lower pressures and more significant as pressure increases. At 2,800 bar, a substantial blue shift happens, followed by the loss of the peak structure above 3,000 bar, which indicates protein unfolding (Fig. 5D). The position of the absorption maximum of PEB chromophore is much less pressure sensitive (Fig. 5E). Indeed, an increase of pressure up to 1,800 bar induces only a modest shift from 552.6 to 553.1 nm. Further increase of pressure up to 2,600 bar induces a first blue shift, while the second blue shift happens above 3,000 bar. Considering the SAXS results and the behaviour of the absorption maximum of PCB chromophore (Fig. 5D), it is unlikely that the first blue shift of PEB chromophore solely arises from the unfolding process but may also include R-PC oligomer dissociation. Moreover, the constant value of PEB absorption maximum from 2,600 to 3,000 bar, followed by the second blue shift above 3,000 bar (Fig. 5E), indicates that these two observed blue shifts correspond to different processes during the R-PC dissociation and unfolding deduced from the SAXS analysis.

The PCB chromophore of R-PC exhibits an intensive fluorescence peak at 637 nm upon excitation at 590 nm at atmospheric pressure (Fig. 6A). The pressure increase substantially reduces the fluorescence intensity. At 3,000 bar, the intensity drops to 1.5% of the initial fluorescence (Fig. 6B). Hence, the fluorescence spectra of R-PC are much more sensitive to pressure than the absorption spectra, as generally observed. Thus, monitoring the pressure dependence of fluorescence emission maximum could give more reliable transition curves. Plotting the pressure dependence of R-PC emission maximum gives a transition curve (Fig. 6C) similar to the absorption study: an increase of pressure up to 1,600 bar induces a moderate redshift to 641 nm, while a further increase of pressure up to 2,600 bar triggers a substantial redshift to 651 nm. Similar to the HP absorption study, further pressure increase, especially above 3,000 bar, induces a blue shift in the fluorescence spectra, confirming that R-PC is unfolding.

The pressure behaviour of PEB fluorescence significantly differs from the behaviour of the PCB emission (Fig. 6D). An increase of pressure up to 1,000 bar induces a slight increase of PEB fluorescence (Fig. 6E) and, from this pressure point, emission intensities are almost constant up to 2,400 bar. The initial increase in PEB fluorescence may be ascribed to the structural rearrangements or dissociation of R-PC oligomers, which may increase the distance between PCB and PEB chromophores. This could lead to decreased FRET phenomenon (Fig. S3) and lower quenching of the donor fluorophore (PEB) by the acceptor (PCB) chromophore. Interestingly, while the absorbance of PEB ligand is significantly reduced at 2,400 bar (Fig. 5C), PEB fluorescence does not significantly decrease up to 2,400 bar (Fig. 6E). Meanwhile, at 2,600 bar and above, a significant decrease in the fluorescence intensity occurs, which arises from further R-PC dissociation and denaturation. Increasing more the pressure, up to 4,000 bar, does not completely quench PEB fluorescence. At the final pressure point (4,000 bar), around 45% of fluorescence intensity is preserved (Fig. 6E).

An increase in pressure triggers the significant red shift of the PEB fluorescence (Fig. 6F). The obtained curve depicts the complex behaviour of R-PC, characterised by several transitions (Fig.

6F). A red shift from 574 to 576 nm is observed up to 2,000 bar, followed by constant values up to 2,400 bar, while the further increase of pressure induces an additional redshift of PEB fluorescence to above 579 nm at 4,000 bar. However, the red shift from 2,400 to 3,000 bar is much less pronounced than the one from 3,000 to 4,000 bar, indicating at least two different processes. While the first red shift relates to the dissociation process in R-PC, the second one may result from both dissociation and unfolding processes.

### 2.3 Reversibility of R-PC structural changes under HP

The drop in pressure from 4,000 to 27 bar (Fig. 7A) induces a slight increase in SAXS intensity and the reappearance of a small peak in the Kratky plot around 0.04 Å$^{-1}$, indicating a weak partial refolding. However, the absence of the characteristic double-peak in the Kratky representation reveals the absence of R-PC trimers. Based on $I$(0) calculations, the obtained molecular weight ($MW$) after depressurisation is around 51 kDa, which roughly corresponds to the average mass of R-PC monomers (αβ) and dimers (αβ)$_2$. This is also confirmed by the GASBOR modelling, which does not give envelopes resembling the trimeric form of R-PC but rather monomeric and dimeric forms (inset in Fig. 7A).

Both VIS absorption and fluorescence studies (Fig. 7, B-D) reveal that a decrease in pressure from 4,000 to 2,000 bar does not induce any significant refolding of R-PC. However, further pressure decrease to 1,000 bar triggers significant changes in R-PC absorption and fluorescence, indicating its refolding. An additional pressure decrease to 1 bar provokes only subtle changes in R-PC absorption and fluorescence.

Absorption and fluorescence peaks of R-PC have similar, but not identical, positions of corresponding maxima compared to the native protein before HP treatment (Fig. 7, B-D). Absorption and fluorescence are partly recovered: 50% and 67% of the initial absorption values and 33% and 74% of the initial fluorescence values for PCB and PEB chromophores, respectively, are measured, meaning that part of the R-PC typical colour is recovered after HP treatment. Interestingly, the decrease of pressure from 1,000 bar to atmospheric pressure induces a slight decrease in PEB fluorescence (Fig. 7D), which is in line with the pressure-induced fluorescence small enhancement up to 1,000 bar (Fig. 6E). It seems that depressurisation from 1,000 to 1 bar induces increased FRET phenomenon between PEB and PCB chromophores, indicating a decreased distance between them and a partial reoligomerization of R-PC. Therefore, depressurisation induces refolding of R-PC, but only partially. This is very different from the thermal treatment of R-PC at 60°C, which induces detrimental and irreversible effects on the R-PC structure with complete loss of the absorption bands (Fig. 7E).

*Ex situ* far-UV circular dichroism (CD) spectroscopy reveals that HP treatment at 4,000 bar irreversibly decreases ellipticity for R-PC (Fig. 7F). Nevertheless, the shape of the CD spectrum of a pressurised sample does not point to a complete unfolding upon HP treatment (Fig. 7F). R-PC pressurisation induces a significant loss of α-helical content, from 49.4 to 19.1%, while β-sheet content substantially increases, from 3.6 to 27.1% upon HP treatment. HP also induces a slight increase in the content of β-turn structure (from 19.7 to 22.2%) and random coil content (from 27.3% to 31.6%).

## 3. DISCUSSION

For the first time, we investigated R-PC *in situ* pressure structural stability by combining SAXS and optical spectroscopies. We discovered the complex behaviour of R-PC under HP: increasing pressure induces conformational changes in the R-PC trimeric structure, followed by trimer dissociation and subunits unfolding. Subsequent pressure release results in partial refolding and oligomerisation, but the native trimers are not fully recovered. Nevertheless, the impact of pressure treatment on R-PC is significantly less detrimental than heat treatment.

The 3D structure of R-PC closely resembles the structure of C-PC from cyanobacteria *Arthrospira platensis*, the most studied PBP. The (αβ) monomers of PBPs have a high tendency to oligomerise into trimers ((αβ)$_3$) and hexamers ((αβ)$_6$), both having specific hollow disc structures (Jiang et al. 2001; Padyana et al. 2001). Our previous SAXS study demonstrated that, at pH 5.7, R-PC is a trimer (Velickovic et al. 2023), unlike C-PC, which oligomerises into hexamers at the same pH value (Saxena 1988). This difference indicates that the presence of the PEB chromophore in R-PC and not in C-PC causes subtle conformational changes that influence the oligomerisation of the protein (Velickovic et al. 2023). It was previously shown that PBP hexamers are much less stable than trimers, and even minor changes in their structure or environment (such as pH, ionic strength, or low concentration of PBP denaturants) can induce dissociation of hexamers into trimers (Marx and Adir 2013). On the opposite, the high affinity of (αβ) monomers for oligomerisation into trimers ($K_a$ between $10^{12}$ and $10^{15}$ M$^{-1}$) makes PBPs in trimeric form the robust system for studying their structural stability under pressure (Thoren et al. 2006).

For most oligomeric proteins, the pressure shifts the equilibrium between oligomers towards dissociation into subunits. This behaviour is due to the inter-subunit regions of oligomeric proteins, rich in electrostatic and hydrophobic contacts that stabilise the quaternary structure and are weakened at relatively moderate pressure (Balny 2002). These inter-subunit regions also contain hydrophobic cavities, which are a critical structural feature and source of fluctuations between different protein conformations, determining their behaviour at HP conditions (de Oliveira et al. 2016; Minic et al. 2020; Winter at al. 2007).

The hollow cylinder shape of R-PC trimers, with characteristic shapes clearly observed in SAXS spectra, makes R-PC a good model system for probing the effects of pressure on oligomeric protein dissociation. A moderate increase of pressure (up to 600 bar) on R-PC induces an immediate decrease in the thickness of the cylinder and of $I(0)$ as well, the last one depending on the protein volume and hydration. These parameters suggest that dissociation of R-PC might occur at low pressure. However, (*i*) the *ab initio* GASBOR model resembling the shape of trimeric R-PC, (*ii*) the well-defined peak at 0.1 Å$^{-1}$ in Kratky plots (an indicator of hollow disc shapes), (*iii*) the good fits of SAXS curves with a hollow cylinder model, and (*iv*) stable $R_g$ values up to 1,500-2,000 bar, confirm the persistence of R-PC trimers up to 1,600 bar. Indeed, the cylinder length and $I(0)$ do not change significantly between 600 and 1,400-1,600 bar, indicating that pressure under 600 bar induces only conformational changes in R-PC trimers.

From 600 to 1,400-1,600 bar, stable intermediates are formed, reflecting a more compact structure with a lower total protein volume. The latter is reasonably responsible for the smaller $I(0)$ values measured rather than a protein dissociation. Another reason for decreased $I(0)$ upon pressure increase up to 600 bar may be the enhanced protein hydration. Although we could not exclude this possibility at moderate pressures, our calculation reveals (data not shown) that the decrease of $I(0)$ of 20%, induced by pressure increase from 30 to 600 bar, would require the binding of 3,000 water molecules *per* R-PC trimer. This is unrealistic, especially if we consider that R-PC remains well folded at 600 bar. Pressure-induced protein hydration may contribute but little to the decreased $I(0)$ values.

GASBOR modelling generates a trimeric-like structure with the central hole as a prominent characteristic up to 1,600 bar, with a significant reduction in the diameter of both the central hole and the disc thickness, consistent with the fitting by a hollow cylinder model. These observations agree with the tightening of the protein globular structure shown for alkaline phosphatase at low pressure (up to 1,500 bar) (Cioni and Strambini 1996). An increase of pressure up to 1,800 bar induces the dissociation of compressed R-PC trimers into structures resembling $(\alpha\beta)_2$ dimers. Although this dimeric form of PBPs is not usually present at native conditions, it seems that HP induces the formation of dimers, as previously shown by pressurisation of C-PC from *Arthrospira platensis* in the range 1,500 to 2,000 bar (Li et al. 2023).

Further pressurisation up to 2,600 bar provokes the dissociation of $(\alpha\beta)_2$ dimers to $(\alpha\beta)$ monomers, as revealed by $I(0)$ values, which are nearly three times smaller than the values of native trimeric R-PC. However, the significant increase of the $R_g$ values of monomers, as well as the inability to fit the experimental SAXS profiles with theoretical curves of the crystal structures of $(\alpha\beta)$ monomers crystal structures (data not shown), mean that the obtained $(\alpha\beta)$ monomers have partially extended shapes as also revealed by GASBOR modelling. Hence, dimer dissociation to monomers is also followed by conformational changes in R-PC subunits, leading to partial unfolding but without dissociation of $(\alpha\beta)$ monomers to individual subunits. An increased pressure to 3,000 bar induces a further increase in $R_g$, followed by an extension of the GASBOR model, suggesting additional protein denaturation. However, the mostly constant $R_g$ values up to 3,400 bar, followed by its additional increase up to 4,000 bar, indicates that R-PC is not entirely unfolded at 3,000 bar, but further pressure increase is required for the protein to unfold completely. The decrease of $I(0)$ from 2,600 to 4,000 bar indicates that the dissociation of $(\alpha\beta)$ monomer to individual subunits is not complete, but rather the unfolded subunits are complexed into non-native $(\alpha\beta)$ monomers. However, we could not exclude the possibility that the complete dissociation and unfolding of α and β subunits precede the formation of non-native $(\alpha\beta)$ monomers, but these processes could not be resolved by this HP-SAXS study.

Covalently attached tetrapyrrole chromophores are valuable indicators of PBP folding and oligomerisation state. The PEB and PCB chromophores of R-PC exhibit different behaviours in response to HP, affecting the intensity and position of their maxima in absorption and fluorescence spectra (Velickovic et al. 2023; Wang et al. 2014). The absorption and fluorescence intensity of the PEB chromophores are much more resistant to pressure changes than those of the PCB chromophores. However, the normalised curves of the PEB and PCB absorption intensities, fitted with the two-state transition model, give similar thermodynamics parameters (volume change, $\Delta G$, and $P_{1/2}$) for R-PC dissociation/unfolding. This indicates no significant difference in the pressure stability of α and β subunits. Therefore, the different dependences on the pressures of PCB and PEB chromophores are not influenced by the stability of individual subunits. Instead, they are influenced by the distinct sensitivity of these tetrapyrrole chromophores. Similarly, the study of the thermal stability of R-PC shows that protein unfolding is a single-step process, although the PEB chromophore, exclusively bound to the β subunit, is more resistant to thermal treatment (Velickovic et al. 2023).

The pressure dependence of PEB fluorescence is highly specific since its emission intensity is more resistant to pressure changes than PEB absorbance. Moreover, a slight increase in PEB fluorescence at low pressure indicates a larger distance between donor (PEB) and acceptor (PCB) chromophores within R-PC trimers. This initial increase of PEB fluorescence (up to 600 bar) corresponds to the formation of a trimeric-like intermediate detected by the hollow cylinder fit of the form factor and by GASBOR modelling. (Ma et al. 2008) demonstrated that a small urea concentration induces subtle conformational changes in R-PC β subunits, increasing the distance between PCB and PEB chromophores. The stable value of PEB fluorescence from 600 to 2,400 bar, despite the significant decrease in PEB absorbance above 1,600 bar, indicates the presence of two opposite effects on the PEB fluorescence: (*i*) the dissociation of R-PC trimers induces the decrease of the emission intensity and (*ii*) the increase of distance between PEB and PCB chromophores during structural rearrangements and dissociation of R-PC trimers contributes to the PEB fluorescence enhancement due to decreased FRET. Hence, these two phenomena cancel each other, leading to constant PEB fluorescence up to 2,400 bar. Beyond this pressure, unfolding (αβ) monomers or individual subunits occur, generating a substantial drop in the PEB quantum yield, which cannot be compensated by the increase in the distance between PEB and PCB in the unfolded subunits.

While simple curves describe the pressure dependence of absorption and emission intensities, the pressure dependence of absorption and fluorescence peak positions yields more informative curves, with several transitions depicting the complex behaviour of R-PC under pressure. Three phases have been observed: (*i*) the slight red shift up to 1600-1800 bar related to the conformational changes within trimeric structures, (*ii*) a steep red shift up to 2,600-2,800 bar due to R-PC oligomer dissociation to (αβ) monomers/individual subunits, and (*iii*) a blue shift above 2,800 bar induced by subunits unfolding. The red shift of tetrapyrrole chromophores is related to the out-of-plane distortion of the pyrrole rings, resulting from the weakened protein-chromophore interactions at HP due to oligomer dissociation (Li et al. 2023). It was demonstrated that the absorption peaks of (αβ) monomers of C-PC and individual subunits are significantly red shifted compared to oligomeric forms (Debreczeny et al. 1993). The pressure-induced dissociation of C-PC trimers is also characterised by the red shift of the PCB absorption peak of *Arthrospira platensis (*Li et al. 2023*)*. Similar differences of red shifts was observed in the renaturation of urea unfolded individual subunits of R-PC compared to the trimeric R-PC (Ma et al. 2008). Here, a modest red shift at lower pressure indicates conformational changes in R-PC trimers, as shown by SAXS, while a more prominent red shift at higher pressure (above 1,800 bar) suggests more extensive oligomer dissociation, as previously described for C-PC (Li et al. 2023). On the other hand, the pressure-induced unfolding of R-PC provokes the changes in PCB chromophore from extended to cyclic shape, leading to the blue shift in absorbance and fluorescence. The urea-induced unfolding of C-PC also induces the shifts of absorption spectra to the same direction (Scheer and Kufer 1977).

Generally, PEB absorption and emission maxima positions are generally less pressure-sensitive than PCB chromophores. Here, PEB absorption maxima experience several small red and blue shifts during pressurisation. On the other hand, the pressure dependence of the PEB emission maxima is characterised by two major phases: (*i*) dissociation of different R-PC oligomeric species (*i.e.* (αβ)$_3$, (αβ)$_2$ or (αβ)) up to 2,400-3,000 bar and (*ii*) unfolding of subunits at and above 3,000 bar. It is worth pointing out that both absorption and fluorescence of the PEB and PCB chromophores demonstrate significant differences between 3,000 and 4,000 bar. This confirms the SAXS results that pressure increases up to 3,000 bar cannot induce complete unfolding of R-PC but rather the formation of the partially unfolded (αβ) intermediate. Moreover, the wavelength dependence of the absorption and fluorescence maxima of PEB reveals a plateau around 3,000 bar, demonstrating the stability of the obtained intermediate.

One of the crucial questions regarding the HP behaviour of proteins under pressure is the reversibility of the pressure-induced unfolding. The gradual decrease of pressure after HP treatment of R-PC solution at 4,000 bar induces partial refolding of the protein without recovering the native trimeric conformation. SAXS reveals that R-PC appears as (αβ) and (αβ)$_2$ oligomerisation forms after HP treatment at 4,000 bar. Conversely, depressurisation of C-PC after treatment at 3,500 bar induces the complete recovery of the protein trimeric structure, as revealed by HP-SAXS (Li et al. 2023). The same study demonstrates that C-PC subunits mostly unfold at this pressure and are not aggregated. It seems that the existence of non-native (αβ) monomers, with mostly unfolded subunits at 3,000 bars and above, followed by additional unfolding up to 4,000 bar, is the critical reason for determining R-PC inability to recover the native oligomeric state after depressurisation. HP absorption and fluorescence studies reveal that depressurisation induces a significant recovery of the position of the absorption and emission peaks (between 70 and 90%) compared to the sample before HP treatment. However, the inability of depressurisation to reach precisely the same value of these peak positions confirms the SAXS-based conclusions of the inability of R-PC to re-associate into trimers after HP treatment. Moreover, depressurisation is less effective in recovering absorption and fluorescence intensity, especially for the PCB chromophore. This can be explained by the pressure-induced oxidation of tetrapyrrole chromophores, leading to the irreversible decrease of absorbance and fluorescence intensities. The PCB chromophore is much more sensitive to oxidation than the PEB molecule (Kohata et al. 2010), leading to the less efficient recovery of the PCB spectroscopic properties after depressurisation. In our previous study, we also demonstrated that depressurisation of R-PE can mostly recover the shape of the absorption spectra without completely recovering the absorption intensities of R-PE tetrapyrrole chromophores (Simovic et al. 2022).

The R-PC trimer, which features a central hole, contains also a large cavity within the α subunit (Fig. 8A) that is located near the PCB chromophore (Jiang et al. 2001). Several salt bridges surround the PCB binding site, with the Arg86-Asp87 bridge being the closest (Fig. 8B). The sensitivity of the PCB chromophore to pressure-induced conformational changes in R-PC can be attributed to the elimination of this cavity and the destruction of surrounding salt bridges under increased pressure.

Interestingly, this large cavity is situated in the α subunit, forming an interface with the β subunit of the adjacent αβ monomer (Fig. 8A). This interface is stabilised by a salt bridge between Arg93 of the α subunit and Asp13 of the β subunit (Fig. 8C). In contrast, the interface between α and β subunits within the same monomer does not feature large cavities, which may explain why the αβ monomers can partially recover upon pressure reduction.

Thus, increased pressure leads to the elimination of the large cavity in α subunits and the disruption of the salt bridges necessary for the association of adjacent αβ monomers. This results in the dissociation of R-PC trimers and subsequent conformational drift of the dissociated monomers/subunits. Consequently, the ability to reassemble into trimers upon pressure decrease is impaired, leading to only partial refolding of R-PC.

HP treatment also induces irreversible changes in the secondary structure content of R-PC, as measured by CD, leading to the transformation of α-helix to the β-sheet secondary structures. Similar results were observed upon pressure treatment of C-PC (Zheng et al. 2020). It was previously shown that changes in the secondary structures of oligomeric proteins can be triggered by dissociation and hydration of the subunit interfaces, inducing the conformational changes and consequently redistribution of secondary structures (Julius et al. 2018). Therefore, rearrangements of the R-PC subunits, including an increase of β-sheet secondary structures, may result from the partial reassociation of non-native (αβ) monomers or individual subunits upon pressure decrease. It is well-known that pressure-induced dissociation of oligomeric proteins can induce the "conformational drift" of protein subunits, which re-associate during the pressure release, generating oligomers with different properties compared to the native ones (Lullien-Pellerin and Balny 2002; Royer at al. 1986). Indeed, the pressure-induced dissociation of trimeric allophycocyanin (APC) at subzero temperatures leads to irreversible changes in its absorption spectra, specifically the inability to recover the peak intensity at 653 nm, which is a crucial spectroscopic feature of the APC trimeric form (Foguel and Weber 1995). The possible mechanism for this hysteresis phenomenon is that the unfolding of R-PC (αβ) monomers at 3,000 bar and above induces irreversible changes in the subunit conformation, disabling the proper renaturation and native trimers recovery after pressure release (Fujisawa et al. 1999).

HP processing involves the application of ultra-HP (1,000 – 6,000 bar) to achieve pasteurisation (Huang et al. 2020). It has an essential advantage in various food treatments, preserving nutrients, colours, and flavours (Simovic et al. 2022). Although HP irreversibly induces some loss of the intensity of the R-PC colour, the thermal treatment is much more detrimental, inducing mostly complete loss of R-PC absorption bands, *i.e.* colour bleaching. Moreover, this colour bleaching is also closely related to tetrapyrrole chromophore oxidation (Minic et al. 2018), leading to the lower bioactive activities of R-PC and consequently losing its comparative advantage as an alternative protein. Therefore, HP treatment of foods and beverages containing R-PC is a promising alternative to thermal processing, particularly avoiding the loss of vivid R-PC colour. However, introducing R-PC as a successful food colourant in the industry requires additional efforts to strengthen its pressure stability. Previous studies demonstrated the ability of sugars as well as whey proteins to increase the pressure and thermal stability of C-PC from cyanobacteria Spirulina, the well-established protein-based food colourant (Faieta et al. 2021; Faieta et al. 2020). Considering the high similarity of the tertiary structures of R-PC and C-PC, this approach deserves to be studied for R-PC. However, the above approach requires a significant concentration of stability-promoting molecules, such as sugars and whey proteins. An alternative approach could be to use food-derived ligands capable of binding specifically to R-PC cavities with high affinity, enabling much higher stabilisation effects at low concentrations. In this respect, it has previously been shown that the high-affinity binding of retinol to the central cavity of bovine lactoglobulin substantially increases the pressure stability of the protein at micromolar concentrations (Minic et al. 2020). Furthermore, the specific binding of food molecules in the central hole of R-PC would substantially stabilise the trimeric structure of R-PC, inhibiting its pressure-induced dissociation, an essential prerequisite for unfolding R-PC subunits.

## 4. CONCLUSIONS

We examined the structural stability under pressure of R-PC, extracted from dried Nori flakes, by combining *in situ* HP-SAXS, HP-VIS absorption, and HP-fluorescence spectroscopy. We investigated the complex behaviour of R-PC during pressure increase, observing compression and flattening of the trimeric R-PC structure, oligomer dissociation into dimers and monomers, extension of monomer conformation, and unfolding. The results obtained by combining HP-SAXS with HP-VIS absorption and HP fluorescence correlate well, confirming that these techniques are invaluable tools for studying the dissociation and unfolding of oligomeric PBPs under HP conditions.

The depressurisation of the R-PC sample from 4,000 bar to nearly atmospheric pressure induces partial renaturation of R-PC. However, this process does not fully restore the native oligomeric state, with high α-helical content, but results in a mixture of structures resembling (αβ) monomers and (αβ)$_2$ dimers with a higher abundance of β sheets. Nevertheless, HP treatment up to 4,000 bar is significantly more effective in preserving R-PC colour intensity than heating at 60°C.

This work elucidates the mechanism of the pressure-induced unfolding of R-PC, a crucial point for developing new approaches to enhance its pressure stability. Further studies are needed to increase the pressure stability of R-PC, potentially by selecting food-derived ligands that may enhance pressure stability and enable proper renaturation, leading to the complete recovery of its colour intensity upon pressure treatment. This would enhance R-PC's potential as a food colourant and bioactive alternative protein.

## 5. MATERIALS & METHODS

### 5.1 Materials

Dried Nori flakes (*Porphyra haitanensis*) were purchased from Fujian Friday Trading Co. Ltd (China). Hydroxyapatite (HA) was bought from BioRad (CA, USA). Diethyl aminoethyl (DEAE) sepharose resin and Superdex Increase 200 10/300 size-exclusion chromatography (SEC) column were obtained from Cytiva (MA, USA). All other chemicals were of analytical reagent grade and Milli-*Q* water was used throughout the experiments. Each experiment (unless otherwise stated) was performed at pH 5.7 (20 mM MES buffer containing 150 mM NaCl).

### 5.2 R-Phycocyanin isolation and purification

R-PC was purified from Nori seaweed leaves (*Porphyra haitanensis*) as described previously (Velickovic et al, 2023) with the addition of SEC as the last purification step to remove traces of R-PE. Experimental details are provided in Supplementary Information (SI).

### 5.3 *In situ* HP visible (VIS) absorbance measurements and analysis

Visible (VIS) absorption spectra under *in situ* HP were recorded on a Cary 3E spectrometer (Varian, Palo Alto, CA, USA) using an HP optical "bomb" (HP cell) with sapphire windows and HP generator, as previously described (Minic et al. 2020) (Simovic et al. 2022). A square quartz cell (with an optical pathlength of 5 mm) containing the sample was positioned within the HP optical "bomb". A parafilm on the top of the cell separated the sample from the pressure-transmitting liquid ($H_2O$). The absorbance of the R-PC solution (0.15 mg/mL) was recorded between 500 and 700 nm, with a bandwidth of 1 nm and a data interval of 0.2 nm, at a scanning speed of 60 nm/min. Spectra were recorded at various pressures (between 0.1 and 4,000 bar, with steps of ~100 or ~200 bar) at 20°C. The pressure was increased at ~100 bar/min, while the equilibration time was set to 3 min at each pressure point. After treating R-PC at 4,000 bar, the pressure was decreased with steps of 1,000 bar at a rate of ~500 bar/min. For each pressure point, the equilibration time was set to 3 min. Measurements of buffer solution were performed at the same conditions as above, and the recorded spectra were subtracted from the corresponding protein spectra. The percentage of protein unfolding/dissociation as a function of the pressure *P* is calculated using the following equation (Minic et al. 2020):

$$\%_{R-PC\ unfolding/disscociation} = \frac{A_{0.1MPa} - A_P}{A_{0.1MPa} - A_{400MPa}} \times 100 \qquad (1)$$

where $A_{0.1MPa}$, $A_P$, and $A_{400MPa}$ represent the maximum absorption intensities of PCB and PEB peaks at 0.1 MPa (1 bar), at the given pressure measured, and at 400 MPa (4,000 bar), respectively. The results are expressed as the percentage of unfolded/dissociated R-PC as a function of pressure. The Gibbs free energy (ΔG at 1 bar) and the apparent volume change of unfolding/dissociation (ΔV) were determined by fitting the pressure denaturation curves with the one-transition model using an equation adapted from (Lange et al. 1996):

$$\%_{R-PC\ unfolding/disscociation} = 100 - \left( \frac{100}{1 + e^{-\frac{\Delta G + \Delta V * P}{RT}}} \right) \qquad (2)$$

The pressure at which one-half (50%) of proteins is unfolded/dissociated represents R-PC half-denaturation/dissociation pressure ($P_{1/2}$).

### 5.4 *In situ* HP fluorescence measurements and analysis

Fluorescence measurements at HP conditions were performed on an SLM 8,000 fluorescence spectrometer (USA) using a HP optical "bomb" (HP cell) with optically flat sapphire windows and a HP generator, as previously described (Minic et al. 2023). A square quartz cell (with an optical pathlength of 5 mm) containing the sample was positioned within the HP optical "bomb", while a parafilm membrane on the top of the cell separated the sample from the pressure-transmitting liquid ($H_2O$). Fluorescence spectra of R-PC solution (0.05 mg/mL) were recorded with a data interval of 0.1 nm and an excitation at a scanning speed of 20 nm/min. PCB and PEB chromophore emission spectra were measured from 620 to 670 nm and from 560 to 590 nm, with the excitation wavelengths at 590 and 488 nm, respectively. Excitation and emission slits were set to 4 and 8 nm, respectively. Spectra were recorded at various pressures between 1 and 4,000 bar, with ~100 or ~200 bar steps. The pressure was increased at a rate of ~100 bar/min. After each pressure change, a delay of 3 min was taken before measurement to reach equilibrium. After treating R-PC at 4,000 bar, the pressure was decreased with steps of 1,000 bar at a rate of ~500 bar/min. For each pressure point, the equilibration time was set to 3 min. The obtained spectra were smoothed using a Savitzky-Golay algorithm with 100 points of the window. The results were expressed as the dependence on pressure of: (*i*) the normalised maximal fluorescence intensity and (*ii*) the position of emission maximum.

### 5.5 HP-SAXS measurements

HP-SAXS measurements were performed on SWING beamline at the French synchrotron facility SOLEIL (St-Aubin, France) (Thureau et al. 2021; Velickovic et al. 2023) using the wavelength $\lambda = 0.764$ Å (16.235 keV) and a sample-to-detector distance of 3 m. The achievable $Q$-range was 0.0037 to 0.34 Å$^{-1}$, where $Q = (4\pi/\lambda) \sin\vartheta$ is the modulus of the momentum transfer and $2\theta$ is the scattering angle. HP-SAXS cell was built as previously published (Skouri-Panet et al. 2006). R-PC (5 mg/mL) and buffer were loaded (70 μL) into the plastic (PEEK) piece containing polyimide (Kapton) windows (Skouri-Panet et al. 2006), while a parafilm membrane on the top of the piece separated the sample from the pressure-transmitting liquid ($H_2O$). The plastic piece was inserted into a HP "bomb" (HP cell) containing diamond windows. The air bubbles in the cell and capillaries were compressed by a slight pressure increase of up to ~30 bar. SAXS spectra were recorded at 20°C at various pressures between 30 and 4,000 bar, with ~100 or ~200 bar steps. The pressure was increased at a rate of ~100 bar/min. After each pressure change, a delay of 3 min was taken before measurement to reach equilibrium. At each pressure step, the sample was measured ten times (frames) for 1 second *per* measurement, with 0.5 second between each frame. After treating R-PC at 4,000 bar, the pressure was decreased with steps of 1,000 bar at a rate of ~500 bar/min. For each pressure point, the equilibration time was set to 3 min.

### 5.6 HP-SAXS analysis

The 2D-SAXS patterns were normalised to the transmitted intensity and azimuthally averaged using the Foxtrot program (Java-based graphical application developed at SOLEIL and available at http://www.synchrotron-soleil.fr/Recherche/LignesLumiere/SWING).

The classical expression of the scattering intensity $I(Q)$ (in cm$^{-1}$) of centrosymmetric and relatively monodisperse particles can be written:

$$I(Q) = n \, \Delta\rho^2 \, V_{\text{part.}}^2 \, P(Q) \, S(Q) \qquad (3)$$

where *n* is the number of particles *per* unit volume (cm$^{-3}$), $\Delta\rho$ (contrast) is the difference between the X-ray SLDs of the particles and of the solvent (cm$^{-2}$), and $V_{part}$ (cm$^3$) is the volume of each particle. The form factor *P(Q)* describes the shape of the particles and fulfils the condition *P*(0) = 1, while the structure factor *S(Q)* represents the interactions between the particles. In the absence of interaction, like in a dilute solution, *S(Q)* = 1.

SAXS data were fitted with a hollow cylinder model (Feigin and Svergun 1987) using the SasView software (http://www.sasview.org). The fitting quality is evaluated using error distribution plots (Fig. S2).

CRYSOL software (Franke et al. 2017) was used to compare the experimental SAXS curve of R-PC in solution at 30 bar with the theoretical curve obtained from the trimeric R-PC crystal structure (PDB:1F99).

We also performed *ab initio* modelling of the R-PC structure, solely based on the SAXS experimental curve, using GASBOR software (ATSAS) (Svergun et al. 2001). The quality of fitting was presented using error distribution plots (Fig. S1). P3 symmetry was used to perform *ab initio* GASBOR modeling of the SAXS curves up to 1,600 bar, while P1 symmetry was used for higher pressures.

The Primus software (Konarev et al. 2003) was used to determine the radius of gyration ($R_g$) of R-PC. This value is measured at small *Q*-values ($QR_g$ in the range 0.4-1.3) using the Guinier approximation:

$$\ln\left(\frac{I(Q)}{I(0)}\right) = -\frac{Q^2 R_g^2}{3} \qquad (4)$$

$R_g$ values were also calculated from the hollow cylinder parameters using the following equation (Feigin and Svergun 1987):

$$R_g^2 = \frac{R_o^2 + R_i^2}{2} + \frac{l^2}{12} \qquad (5)$$

where $R_o$, $R_i$, and $l$ represent the total radius, the inner radius (*i.e.* the radius of the hole), and the length of the cylinder, respectively.

**5.7 *Ex situ* circular dichroism (CD) spectroscopy measurements**

Far-UV CD measurements of R-PC before and after HP treatment at 4,000 bar (400 MPa) were carried out on a Chirascan Plus spectropolarimeter (Chirascan, UK) under constant nitrogen flow. All spectra were recorded at 25°C. Far-UV CD spectra of 100 μg/mL R-PC solution were recorded in the 200–260 nm range at a scan speed of 12 nm/min, using a cell with an optical pathlength of 1 mm and accumulating two scans. The secondary structure content was determined using the COUNTIN algorithm and the SP29 database (Sreerama and Woody 2000).

**5.8 Effect of heating on R-PC absorption properties**

The R-PC solution (0.15 mg/mL) was incubated at 60°C for 5 min. After cooling down to room temperature, the absorption spectrum was recorded on an Agilent Cary 100 Bio spectrophotometer (Agilent, USA) in the range from 500 to 700 nm. The spectrum of the protein solution before heating was used as a control.

**5.9. Visualisation of R-PC structure**

The crystal structure of the trimeric R-PC (PDB:1F99) was visualised using PyMOL software. To identify cavities within the R-PC trimer, the cavity detection radius was set to 3 solvent radii, and the cavity detection cut-off was set to 5 solvent radii. The maximal distance for the formation of salt bridges was set to 4 Å.

## SUPPLEMENTARY MATERIAL

This section includes six figures: **1**) Quality of fitting R-PC SAXS curves with the ab initio models obtained by GASBOR software (Fig. S1), **2**) Quality of fitting R-PC SAXS curves with the hollow cylinder analytical model in SasView software (Fig. S2) and **3**) The overlap in the absorption and emission spectra of R-PC to demonstrate the possible FRET phenomenon between PEB and PCB chromophores (Fig. S3). The supplementary Material section also includes the protocol for the R-PC purification.

## AUTHOR CONTRIBUTIONS

**Simeon Minic:** Conceptualisation, Investigation, Formal analysis, Funding acquisition, Writing – original draft, Writing – review and editing. **Luka Velickovic:** Investigation. **Burkhard Annighöfer:** Investigation. **Aurélien Thureau:** Investigation. **Nikola Gligorijevic**: Investigation, Writing – review and editing. **Zorana Jovanovic:** Investigation. **Annie Brûlet**: Investigation, Formal analysis, Writing – review and editing. **Sophie Combet:** Conceptualisation, Investigation, Formal analysis, Funding acquisition, Writing – review and editing.

## ACKNOWLEDGMENTS

We thank the SOLEIL synchrotron for SAXS beamtime (project No. 20220590) and SWING beamline staff for assistance. This work was supported by 1) the Alliance of International Science Organizations (ANSO), Project No. ANSO-CR-PP-2021-01; 2) Ministry of Science, Technological Development and Innovation of Republic of Serbia contracts No. 451-03-66/2024-03/200168 and 451-03-66/2024-03/200026.; and 3) a short-term FEBS fellowship to Simeon Minic.

**CONFILCT OF INTEREST**

No potential conflict of interest was reported by the authors.

**TABLES**

**Table 1.** Thermodynamic parameters of pressure behaviour of R-PC (0.15 g/L, pH 5.7) obtained from the analysis of the HP absorption intensities of PCB and PEB chromophores.

| Chromophore | $\Delta V$ (mL/mol) | $P_{1/2}$ (bar) | $\Delta G_{0.1\ \text{MPa}}$ (kJ/mol) |
|---|---|---|---|
| PCB | $-92 \pm 1$ | $2{,}330 \pm 3$ | $21.4 \pm 0.2$ |
| PEB | $-84 \pm 1$ | $2{,}450 \pm 5$ | $20.5 \pm 0.3$ |

**FIGURE CAPTIONS**

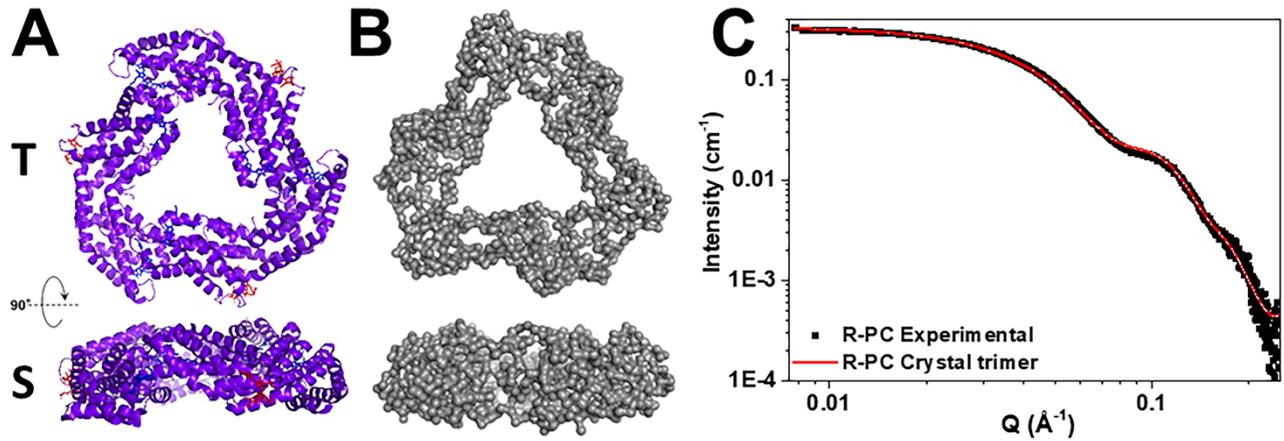

**Figure 1.** Comparison of (*A*) the crystal structure (PDB:1F99) and (*B*) *ab initio* GASBOR (ATSAS) model of trimeric R-PC. (*C*) Superposition of the experimental SAXS curve of R-PC at 5 mg/mL and 30 bar, pH 5.7 (black squares) and of the theoretical SAXS curve of R-PC crystal structure calculated by CRYSOL software (ATSAS, red line).

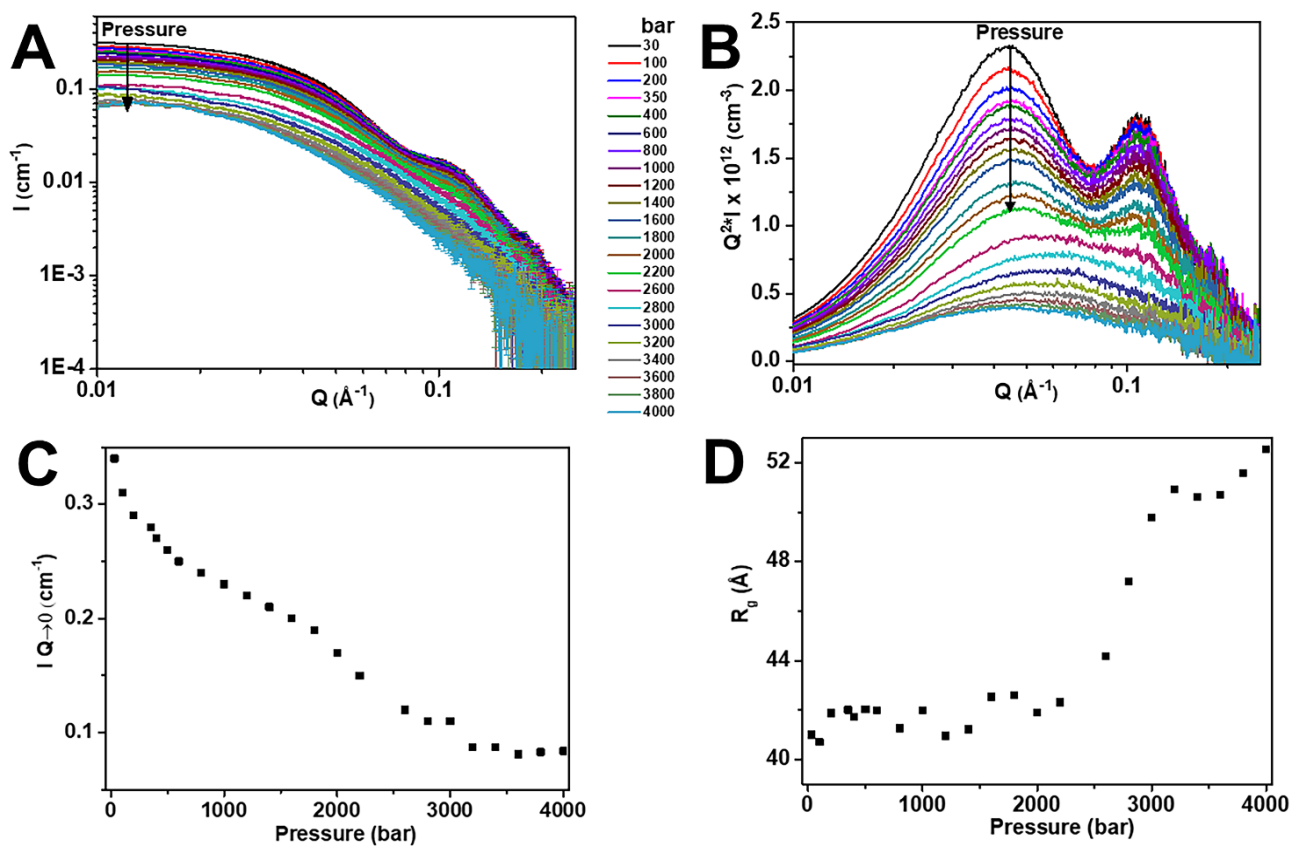

**Figure 2.** (*A*) *In situ* HP-SAXS intensities of R-PC solution (5 mg/mL, pH 5.7) at different pressures from 30 to 4,000 bar (3-400 MPa). (*B*) The same data shown in the Kratky representation $Q^2 I(Q)$ versus $Q$. Evolution of (*C*) $I(0)$ and (*D*) $R_g$ as a function of increasing pressure.

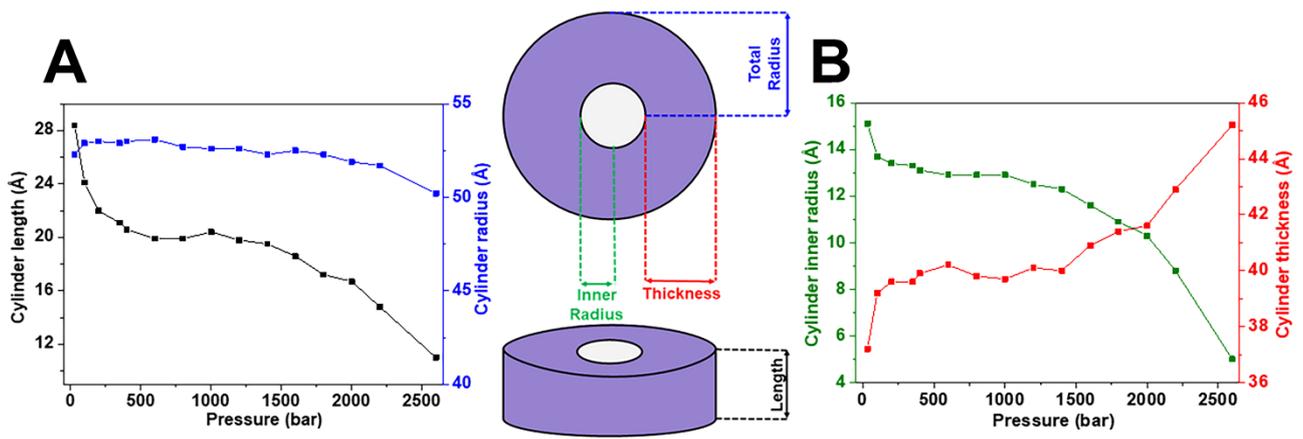

**Figure 3.** Pressure dependence of parameters of the hollow cylinder model: (*A*) length (black) and total radius (blue) and (*B*) inner radius (*i.e.* radius of the cylinder central hole, green) and thickness (red), obtained by fitting HP-SAXS data of R-PC solution (5 mg/mL, pH 5.7) under pressure.

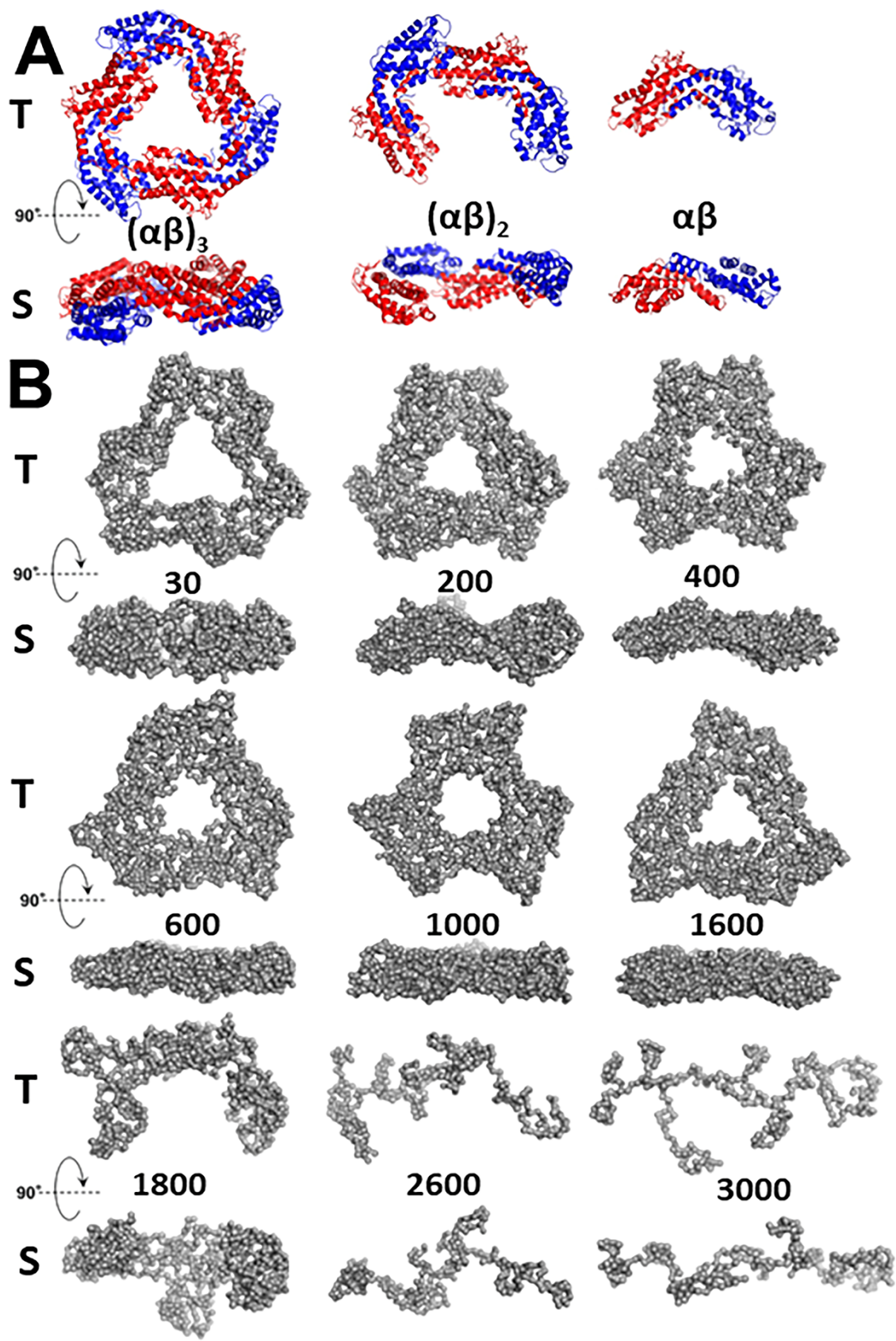

**Figure 4.** (*A*) The ribbon model of the crystal structure of R-PC (PDB:1F99), presented in different oligomerisation states. The R-PC subunits, α and β, are presented in blue and red, respectively. (*B*)

*Ab initio* envelopes of the R-PC oligomers at different pressures (in bar), corresponding to the GASBOR models (ATSAS). Top (T) and side (S) views are shown for each structure.

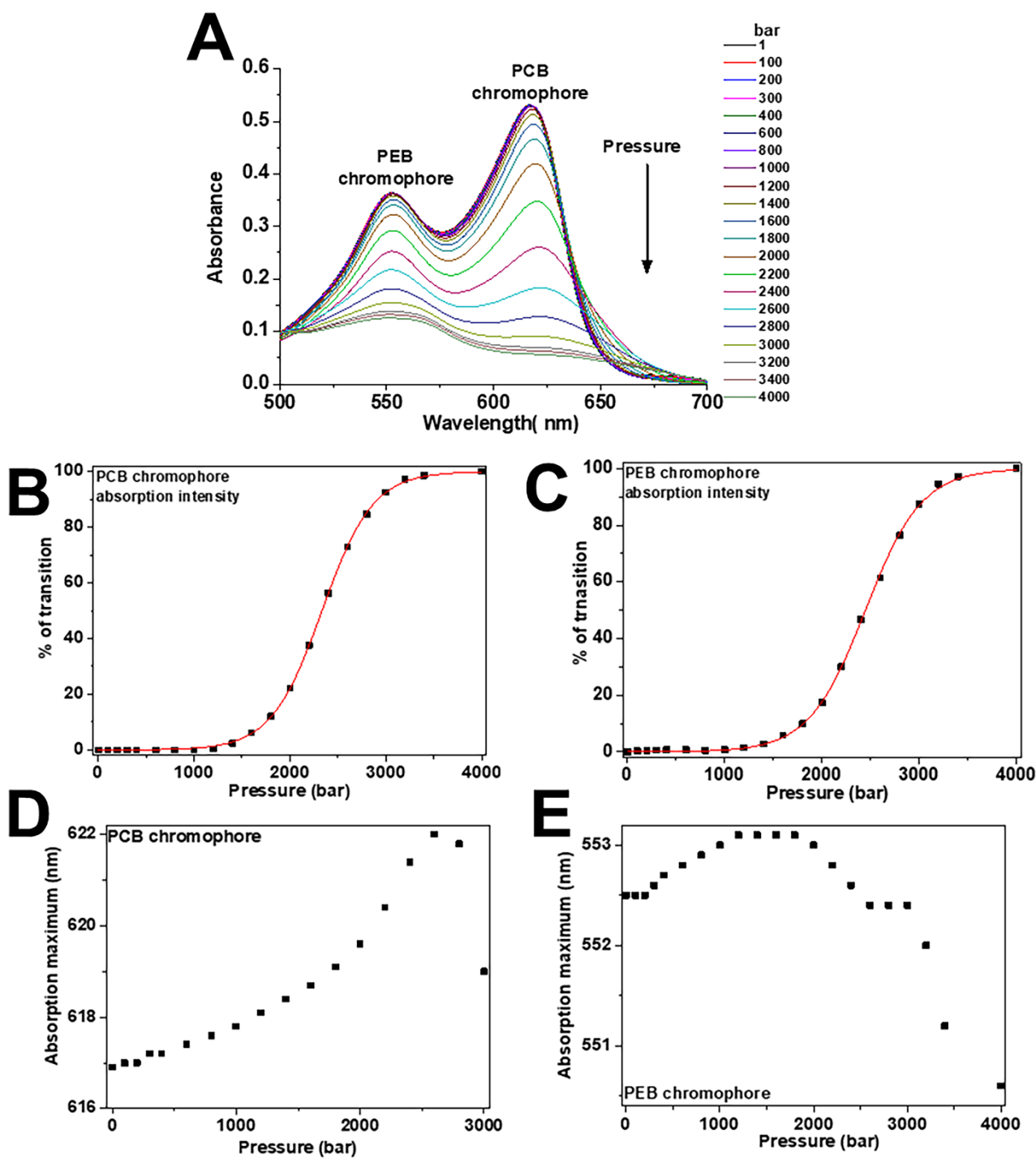

**Figure 5.** (*A*) Effect of HP on the VIS absorption spectra of R-PC. R-PC dissociation/unfolding curves were obtained by plotting the pressure dependence of normalised absorption intensities (*i.e.* percentage of transition) for (*B*) PCB (intensity at the peak at ~617 nm) and (*C*) PEB (intensity at the peak at ~552 nm) chromophores, calculated by Eq. 1. Obtained curves are fitted using Eq. 2 (full red

lines) to determine the thermodynamic parameters of unfolding/dissociation. Effect of HP on the shift of the absorption peaks of (*D*) PCB and (*E*) PEB chromophores of R-PC.

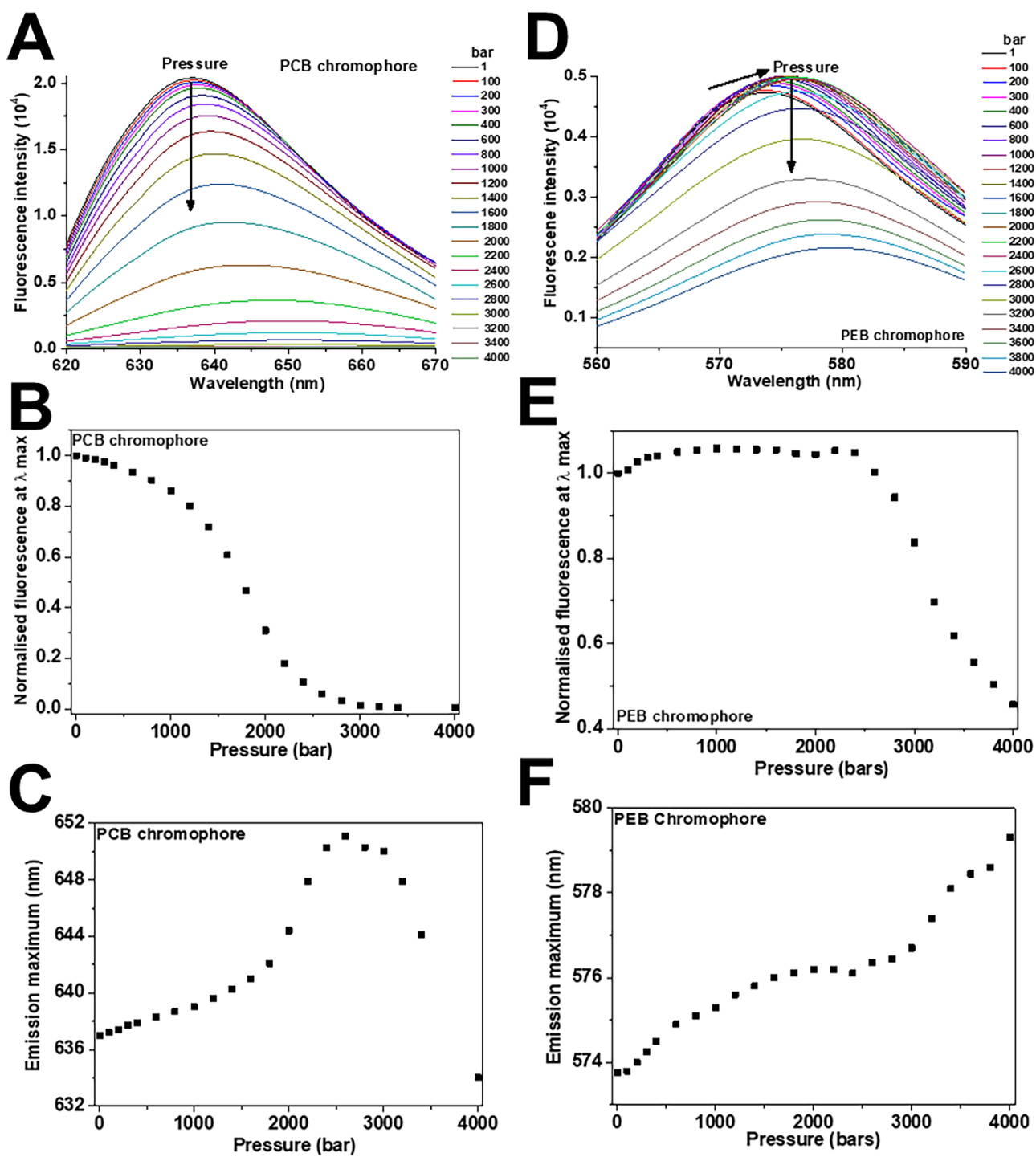

**Figure 6.** Effect of HP on the fluorescence emission spectra of (*A*) PCB and (*D*) PEB chromophores in R-PC, obtained by excitation at 590 and 488 nm, respectively. Pressure dependence of (*B*) PCB and (*E*) PEB fluorescence intensity. Effect of the pressure increase on the shift of the fluorescence emission peaks of (*C*) PCB and (*F*) PEB chromophores.

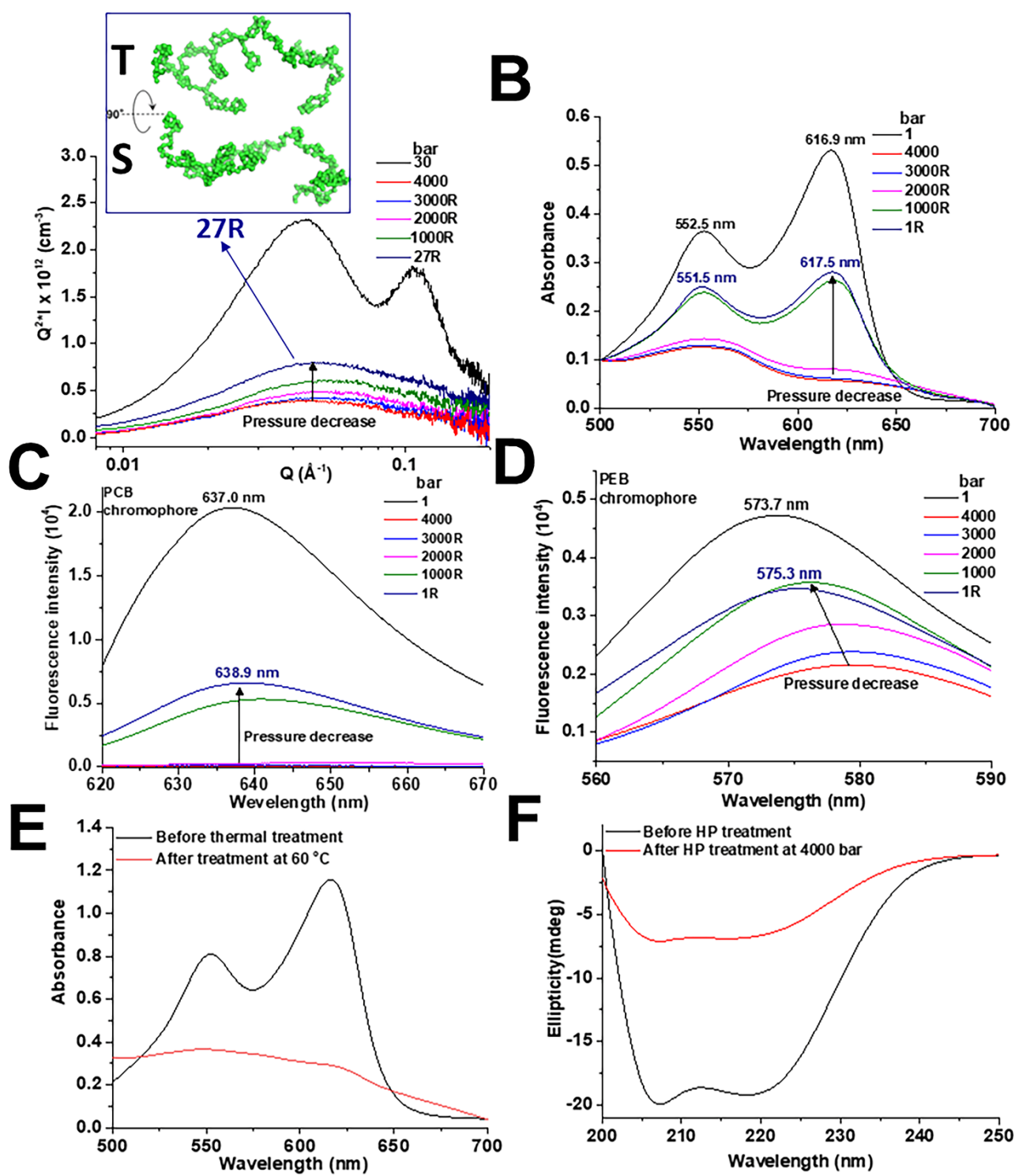

**Figure 7**. (*A*) Effect of depressurisation from 4,000 bar on the SAXS curves of R-PC in the Kratky representation (R in the legend of the curves denotes depressurisation measurements). *Inset:* GASBOR *ab initio* envelope obtained by modeling the SAXS curve recorded after pressure decrease from 4,000 bar to a value close to the atmospheric pressure (27 bar). (*B*) Effect of pressure decrease

on the absorption spectra of R-PC. Effect of pressure decrease on the fluorescence spectra for (*C*) PCB and (*D*) PEB chromophores upon protein excitation at 488 and 590 nm, respectively. (*E*) Effect of temperature on the absorption spectrum of R-PC at pH 5.7 and concentration of 0.15 mg/mL. (*F*) *Ex-situ* effect of pressure at 4,000 bar on the far-UV CD spectra of R-PC.

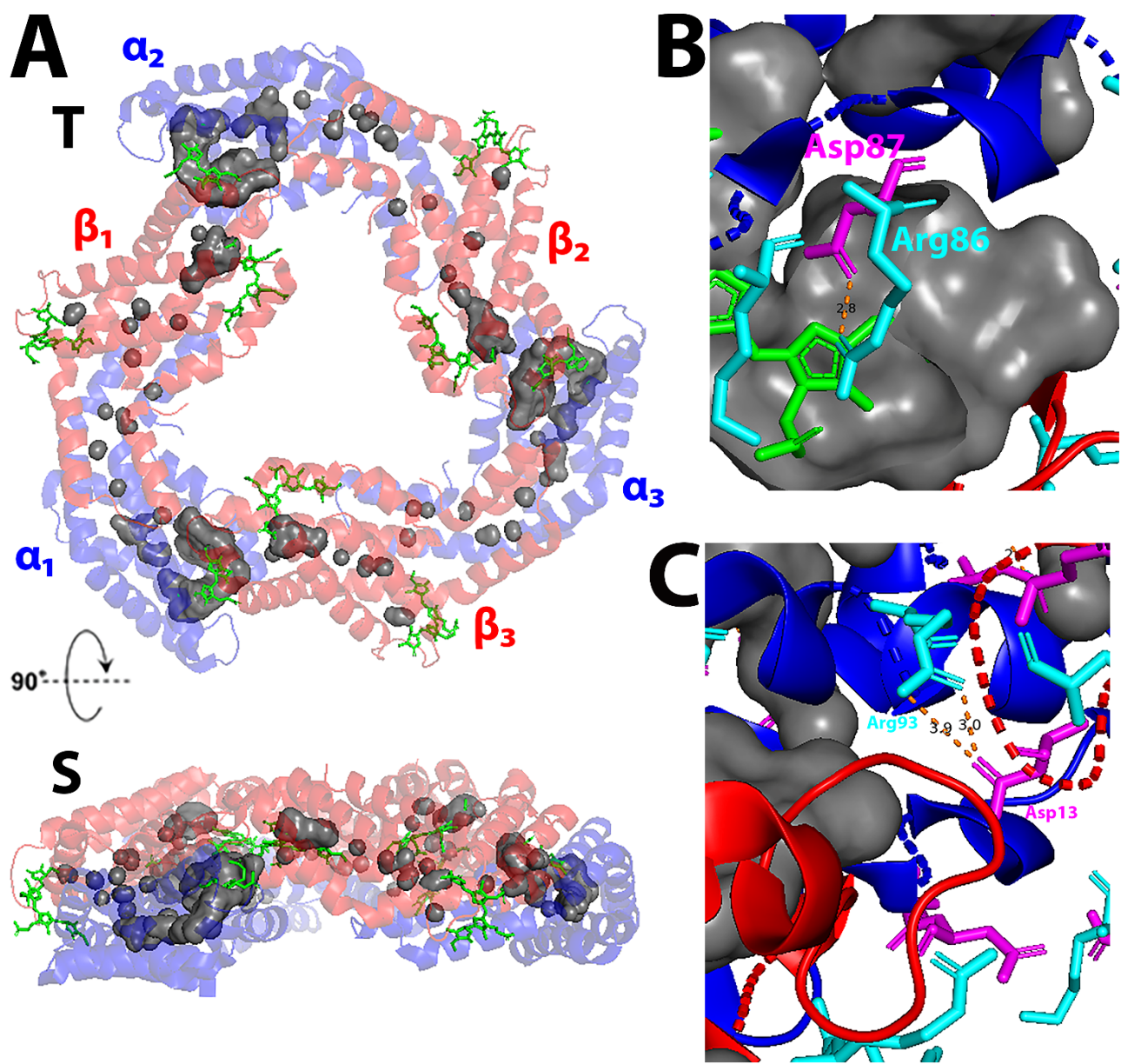

**Figure 8.** (*A*) The ribbon model of the crystal structure of R-PC (PDB:1F99), visualised using PyMOL software. The cavities are shown in grey, while the R-PC subunits, α and β, are coloured in blue and red, respectively. Subunits with the same subscript number belong to the same αβ monomer. The tetrapyrrole chromophores are depicted as a green stick models. The views are labeled as T (top) and S (side); (*B*) Representation of the salt bridge (orange dashed line, 2.8 Å distance) in the vicinity of the grey cavity in the α subunit. This bridge is formed between the Arg86 (cyan stick model) and Asp87 (magenta stick model) residues from the same α subunit; (*C*) Representation of the salt bridges

(orange dashed lines, 3.0 and 3.9 Å distances) formed between the Arg93 (from α subunit, in blue) and Asp13 (from β subunit, in red) residues. The α and β subunits belong to different αβ monomers.